# Predicting the future applications of any stoichiometric inorganic material through learning from past literature

Yu Wu[a], Teng Liu[a], Haiyang Song[a], Yinghe Zhao[a, *], Jinxing Gu[b], Kailang Liu[a], Huiqiao Li[a], Jinlan Wang[c, *] and Tianyou Zhai[a, *] 1

[a]State Key Laboratory of Materials Processing and Die & Mould Technology, School of Materials Science and Engineering, Huazhong University of Science and Technology, Wuhan 430074, China
[b]Department of Chemical & Biological Engineering, Monash University, Clayton Campus, Melbourne 3800, Australia
[c]School of Physics, Southeast University, Nanjing 211189, China

**Abstract**

Through learning from past literature, artificial intelligence models have been able to predict the future applications of various stoichiometric inorganic materials in a variety of subfields of materials science. This capacity offers exciting opportunities for boosting the research and development (R&D) of new functional materials. Unfortunately, the previous models can only provide the prediction for existing materials in past literature, but cannot predict the applications of new materials. Here, we construct a model that can predict the applications of any stoichiometric inorganic material (regardless of whether it is a new material). Historical validation confirms the high reliability of our model. Key to our model is that it allows the generation of the word embedding of any stoichiometric inorganic material, which cannot be achieved by the previous models. This work constructs a powerful model, which can predict the future applications of any stoichiometric inorganic material using only a laptop, potentially revolutionizing the R&D paradigm for new functional materials

## 1. Introduction

The research and development (R&D) of new functional materials is one of the most important topics in chemistry of materials. Each year, a large number of new stoichiometric inorganic materials (SIMs) are synthesized. Exploring their functional applications by traditional trial-and-error approaches, however, is often laborious, cumbersome, expensive, and time-consuming, which significantly increases the time and cost for the R&D of new functional materials. Prior to trial-and-error processes, predicting the promising applications of new SIMs substantially reduces the R&D time and cost. The rise of artificial intelligence opens unprecedented opportunities for accelerating the R&D [1–15]. Although many artificial intelligence (AI) models have been constructed, they are designed for a particular system and a particular application.

Materials science literature records a vast body of valuable knowledge related to materials science [16,17]. To unlock the potential of the knowledge in materials R&D, Tshitoyan et al. constructed a universal AI model for predicting the applications of SIMs [18]. More specifically, the constructed model is the first AI model capable of predicting the applications of various SIMs in a variety of subfields of materials science. Their approach for model construction comprises two main steps: (i) collection of a vast body of literature and (ii) encoding of words extracted from the collected literature into embeddings.

However, a serious challenge facing their model is that it cannot predict the applications of SIMs that do not exist in the collected literature [19]. In other words, the model constructed via their approach can only find the unrecognized applications of existing SIMs in the collected literature [19]. Actually, a large number of new SIMs—they do not exist in any available literature published previously—are synthesized each year. Accordingly, their model cannot predict the applications of these new SIMs. An AI model having the advantage of university and capable of predicting the applications of new SIMs is clearly of greater importance for the R&D of new functional materials, but has not yet been reported.

* Corresponding author. Tel.: +0-000-000-0000 ; fax: +0-000-000-0000 .
*E-mail address:* Zhao, Y. H. (zhaoyh@hust.edu.cn), Wang, J. L. (jlwang@seu.edu.cn), Zhai, T. Y. (zhaity@hust.edu.cn)

In this work, we successfully construct such a model. Namely, our model can predict the applications of any SIM (regardless of whether it is an existing material or a new material) in a variety of subfields of materials science. We further demonstrate the high reliability of our model by historical validation of predictions (HVP). Our model is on par with the model constructed via the approach of Tshitoyan et al. in predicting the applications of existing SIMs. More importantly, our model also offers impressive performance in predicting the applications of new SIMs (the performance even exceeds that in the prediction for existing SIMs). Finally, we showcase the fascinating prospect of our model for the R&D of new functional materials through new SIMs extracted from the journal of Advanced Materials.

## 2. Results and Discussion

*2.1. Model construction*

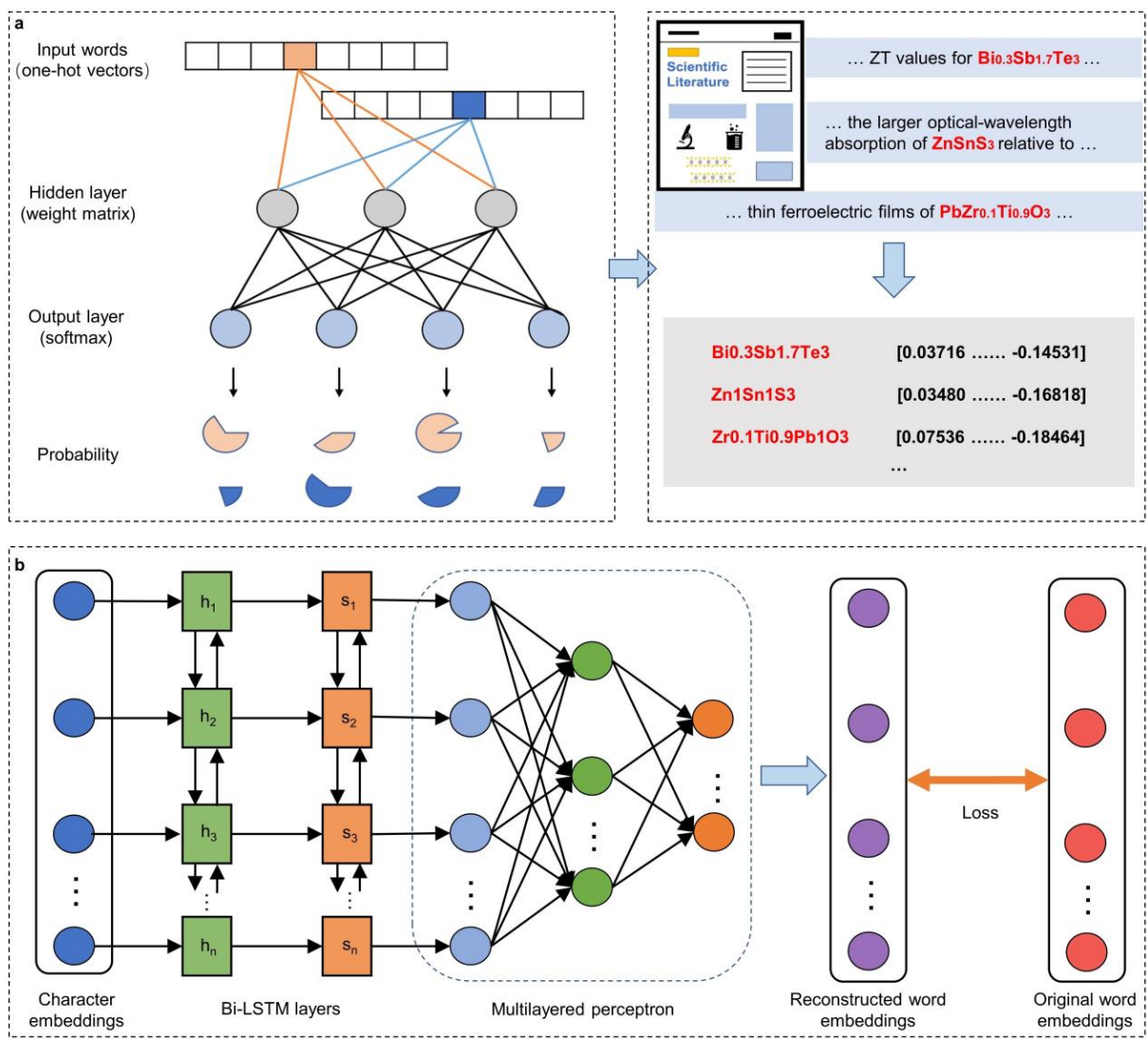

Fig. 1. Schematic diagram for the generation of RWEs. (a) Generation of word embeddings by Word2vec skip-gram. The word embeddings of words in the established corpus are generated by four main steps: (i) the conversion of input words into recognizable one-hot vectors, (ii) the introduction of weight matrix to lower the dimensionality of one-hot vectors in the single linear hidden layer, (iii) the calculation of normalized probability for predicting the context words of each word in the corpus by softmax function in the output layer, and (iv) the continuous update of the hidden-layer weight matrix. The rows of the final hidden-layer weight matrix are the desired word embeddings. (b) Generation of RWEs by MIMICK. More details about the generation of RWEs can be found in the "Methods" section.

The first two steps of the model construction via the previous approach [18] are the same as those via our approach, namely literature collection and corpus establishment. A vast body of literature published from a particular year and earlier needs to be collected to establish a corpus. Notably, the selection of the year is arbitrary, as long as the collected literature is enough. Establishment of the corpus is based on the abstracts and titles of the collected literature. More details about the two steps can be found in the "Methods" section. The next step for model construction is that words in the corpus are encoded into embeddings. Word embeddings (vector representations of words), a fundamental concept in natural language processing, represent the mapped high-dimensional vectors converted from words in the corpus. Word embeddings well preserve semantic relationships underlying the literature [20,21], and more importantly, the relationships of two words can be estimated reasonably by calculating cosine similarity between their word embeddings. The calculation details on cosine similarity can be found in Supplementary Material §1 and Fig. S1. Accordingly, the capacity of word embeddings to characterize the relationships of two words can be used to capture the relationships between SIMs and target applications, making it possible to predict the applications of SIMs.

The previous approach [18], which depends only on Word2vec [22,23], however, cannot generate the embeddings of SIMs that do not exist in the corpus and thus fails to provide the prediction for them. Our approach is based on the combination of Word2vec and MIMICK [24]. MIMICK has the capacity to reconstruct word embeddings. It can capture the composition characteristics of a chemical formula in sequence. For instance, the formula "$CrI_3$" is first recognized as "Cr1I3", and via MIMICK, "Cr1I3" is then split into a set of "C", "r", "1", "I", and "3". Each split character is assigned a character embedding, and then all the character embeddings are combined and converted into the corresponding word embedding. Therefore, the fully trained MIMICK can generate the embedding of any desired SIM, which allows the prediction for both existing and new SIMs. The key to our model is the reconstruction of the word embeddings. The flow diagram for generating the reconstructed word embeddings (RWEs) is shown schematically in Fig. 1. The first step for generating RWEs is the preparation of a training set for MIMICK. The training set consists of chemical formulae in the established corpus and their original word embeddings, which are generated by Word2vec (Fig. 1a). MIMICK is then trained, as shown in Fig. 1b. The training process is controlled by a loss function used to measure the difference between RWEs and original word embeddings.

*2.2. Prediction for the future applications of existing SIMs*

HVP at some point in time refers to the accuracy analysis of the prediction by means of subsequent literature. We used the model that was constructed based on the collected literature from 2001 and earlier as an example to illustrate the meaning of HVP. The model can be used to select promising SIMs for a target application. Promising SIMs refer to SIMs that have never been reported to serve the target application in the literature from 2001 and earlier. The HVP result by a particular year refers to the percentage of those subsequently reported SIMs (SIMs that are subsequently reported to serve the target application between 2002 and the particular year) in the selected promising SIMs. The previous study has shown that HVP is a very effective method for substantiating the reliability of the model in predicting the future applications of SIMs [18].

While RWEs were used in our model, HVP shows that the performance of our model is comparable to that of the state-of-the-art model [18] in predicting the future applications of existing SIMs. Three applications— "thermoelectric", "dielectric", and "photovoltaics"—were chosen as validation cases, for ease of comparison with the previous model. Four corpora were established based on the abstracts and titles of the published literature during four time periods: 2001 and earlier, 2006 and earlier, 2011 and earlier, and 2016 and earlier, respectively; accordingly, based on the above corpora, four models were constructed, respectively (denoted as Models I, II, III, and IV). In keeping with the previous study [18], the chemical formulae of SIMs in the corpus overlapping with those stored in the Materials Project were extracted for HVP. Accordingly, for the years 2001, 2006, 2011, and 2016, four sets consisting of 6301, 7244, 8173, and 9029 SIMs were established, respectively (denoted as Sets 1, 2, 3, and 4). Next, predictions were conducted using the constructed models. Specifically, each model selected the top 50 promising SIMs from the corresponding set for the three target applications, respectively. HVP was performed for the selected SIMs (Fig. 2a–c). Overall, the prediction accuracies gradually increase with the increase of the subsequent year. Importantly, our model shows comparable prediction accuracies with the previous model. Taking the prediction made by Model I for example, the prediction accuracies by the year 2019 for thermoelectric, dielectric, and photovoltaic applications are 46%, 28%, and 44%, respectively, and their sum is 118%. The corresponding accuracies obtained from the previous model are approximately 30%, 26%, and 62%, respectively, and their sum is also 118% (Fig. S2). This shows that a comparable performance is achieved when using RWEs to predict the future applications of existing SIMs in the corpus.

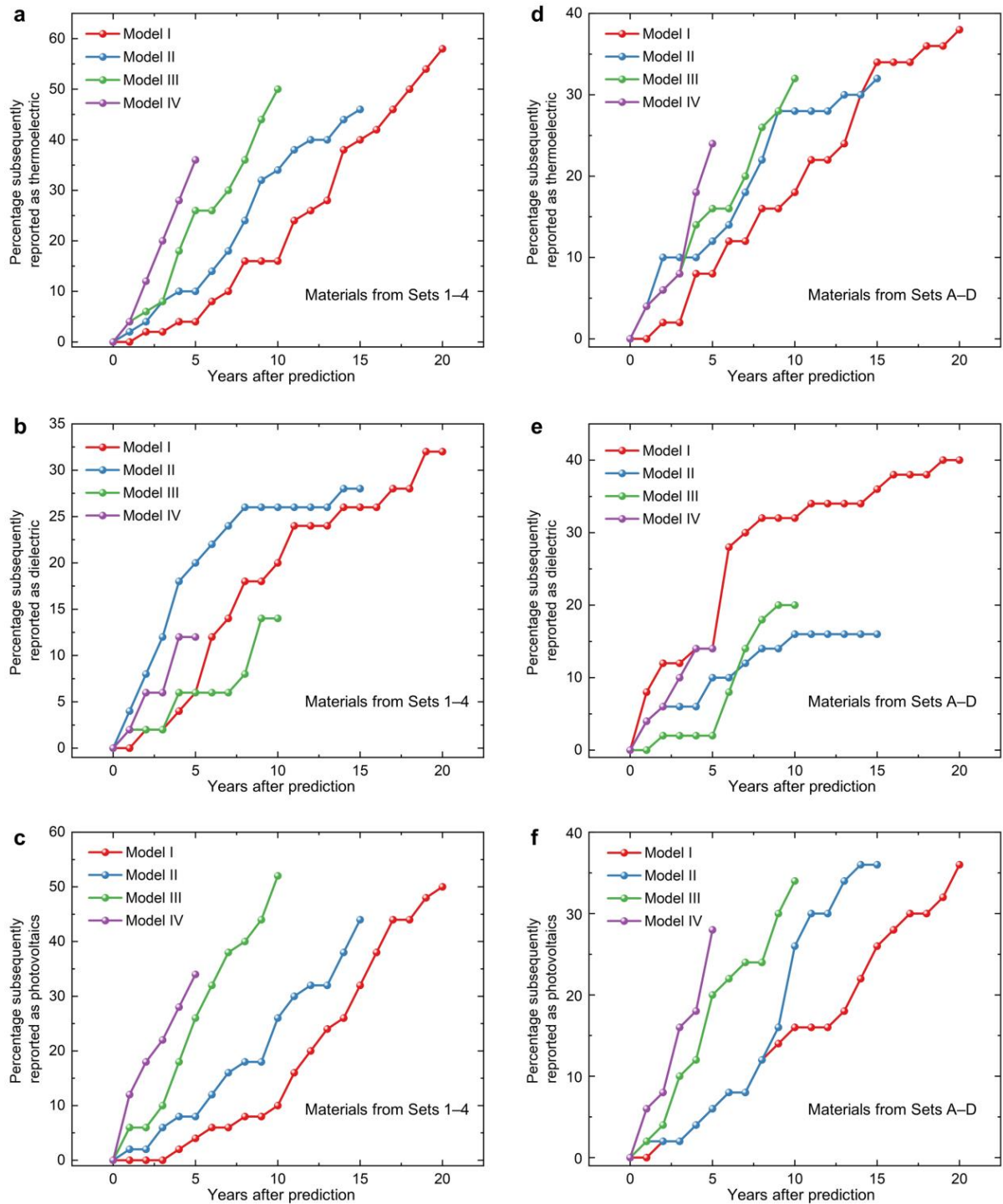

Fig. 2. HVP for existing SIMs. (a–f) HVP results of selected materials for three target applications: (a and d) "thermoelectric", (b and e) "dielectric", and (c and f) "photovoltaics". Red, blue, green, and purple lines record the HVP results of SIMs that were selected by Models I, II, III, and IV, respectively. For each target application, each model selected the 50 most promising SIMs from the corresponding materials set. In (a–c), the corresponding sets for Models I–IV are Sets 1–4, respectively. In (d–f), the materials sets for the selection of Models I–IV are Sets A–D, respectively. Since the materials in Set 1 and Set A exist in the literature used to construct Model I, they belong to existing materials for Model I. Similarly, the materials in Sets 2–4 and Sets B–D also belong to existing materials for Models II–IV, respectively.

In addition, we explored the performance of the model in predicting the applications of all existing SIMs in materials science literature—so as to compare with its performance for the prediction of the applications of new SIMs. All chemical formulae of SIMs in the four corpora mentioned above were directly extracted. The numbers of SIMs in the corresponding sets are 73049, 100100, 133132, and 163326, respectively (denoted as Sets A, B, C, and D, respectively). Each model selected the top 50 promising SIMs from the corresponding set for thermoelectric, dielectric, and photovoltaic applications, respectively, and HVP was then perform for the selected SIMs (see Fig. 2d–f for the HVP results). A direct comparison between the HVP results for existing and new SIMs is presented in the next section.

*2.3. Prediction for the future applications of new SIMs*

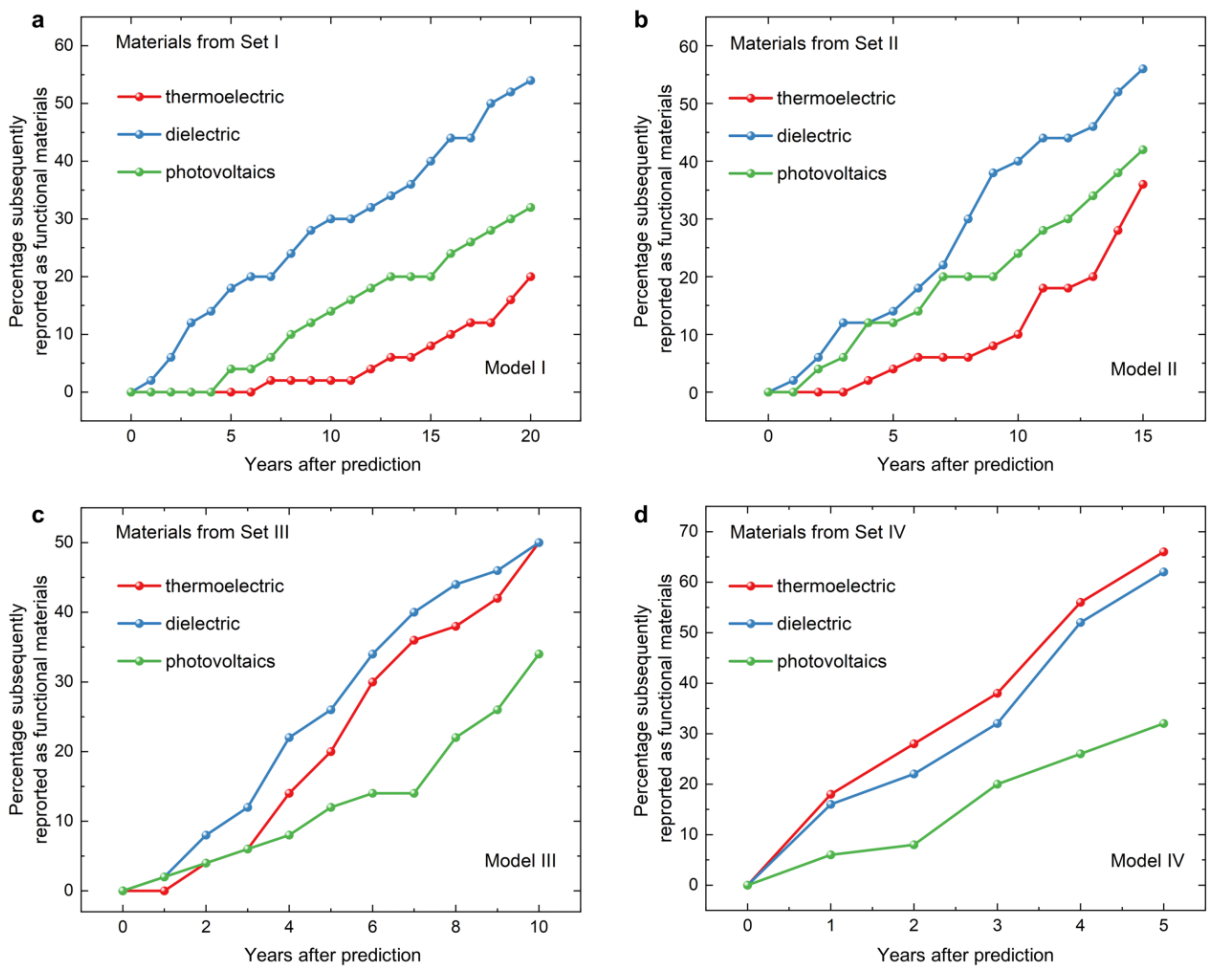

Fig. 3. HVP for new SIMs. (a–d) HVP results of SIMs that were selected from (a) Sets I, (b) II, (c) III, and (d) IV by Models (a) I, (b) II, (c) III, and (d) IV. Each model selected the 50 most promising SIMs from the corresponding set for thermoelectric (red), dielectric (blue), and photovoltaic (green) applications, respectively. Models I, II, III, and IV were constructed based on the literature from 2001 and earlier, from 2006 and earlier, from 2011 and earlier, and from 2016 and earlier, respectively, while Sets I, II, III, and IV were established based on the literature after the year 2001, 2006, 2011, and 2016, respectively. Therefore, the materials in Sets I, II, III, and IV belong to new materials for Models I, II, III, and IV, respectively.

Published literature from a particular year and earlier was collected, and then the collected literature was used to construct the model. Since the constructed model has never seen SIMs that do not exist in the collected literature, SIMs beyond the collected literature belong to new materials for the constructed model. Each year, many SIMs that do not exist in any available literature are synthesized. These newly emerging SIMs are necessarily new materials for the constructed model, regardless of the particular year for literature collection. Although the model constructed via the previous approach has the advantage of university for predicting the applications of SIMs, it cannot provide any prediction for materials beyond the collected literature (i.e., new materials for the model) [19]. To predict the future applications of these newly emerging SIMs, a model having the advantage of university and capable of predicting the applications of new SIMs is required. Remarkably, the model constructed via our approach holds this capacity.

To validate the capacity of our model for predicting the applications of new SIMs, sets of new SIMs need to be established. For some point in time, materials that were reported in the literature after the time point but had never been documented in the literature before the point are collected as new materials. For instance, a material that was reported after the year 2011 but had never been mentioned in the literature from 2011 and earlier is regarded as a new material for the year 2011. The models used here are the same as those used in the above section. Accordingly, four sets of new SIMs were established (denoted as Sets I, II, III, and IV, respectively), and they were extracted from the titles and abstracts of the published literature during four time periods: the years 2002–2022, 2007–2022, 2012–2022, and 2017–2022, respectively. It is worth emphasizing that the corpus for model construction includes no materials in the corresponding set. For example, for Model I, the corresponding set of new SIMs (i.e., Set I) was established based on the literature between 2002 and 2022, while Model I was constructed based on the literature from 2001 and earlier. In other words, the constructed model has never seen those materials in the corresponding set, and they are completely new materials for the constructed model.

Next, HVP was conducted to quantitatively assess the performance of our model in predicting the future applications of new SIMs (Fig. 3). SIMs that are subsequently reported for the target applications accumulate over time, in line with the HVP results for predicting the applications of existing SIMs (Fig. 2). The best model for predicting thermoelectric and dielectric materials is Model IV. Accuracies of 66% and 62% are achieved through the model. Model II performs best for the prediction of photovoltaic materials, and it achieves an accuracy of 42%. The highest prediction accuracies for the three target applications of existing SIMs are 38%, 40%, and 36%, respectively (Fig. 2d–f). The sums of the highest prediction accuracies for existing and new SIMs are 114% and 170%, respectively. Therefore, performance quantitatively superior to that in predicting the future applications of existing SIMs is achieved.

Finally, we focus on showing the fascinating prospect of our model for the future R&D of new functional materials. We constructed two models based on the abstracts and titles of the literature from 2021 and earlier, which were constructed via the previous approach [18] and our approach, respectively. A total of 21 new SIMs were extracted from the abstracts and titles of Advanced Materials published between January 2022 and December 2022 (see Table S1 for the list of materials and the corresponding DOI numbers). The construction of the two models is based on the abstracts and titles of the literature from 2021 and earlier, and the 21 materials have never been mentioned in the abstracts and titles of the literature from 2021 and earlier. Therefore, the 21 materials belong to new materials for the two models. A total of 14 target applications, which are currently the subjects of intense research in materials science, were considered here. They can be divided into two categories: first, physics-related applications, namely "dielectric",, "magnetic", "NLO" (nonlinear optical), "photovoltaic", "superconductors", "thermoelectric", and "TIs" (topological insulators); second, chemistry-related applications, namely "LIBs" (lithium-ion batteries), "HER" (hydrogen evolution reaction), "OER" (oxygen evolution reaction), "ORR" (oxygen reduction reaction), "PIBs" (potassium-ion batteries), "SIBs" (sodium-ion batteries), and "ZIBs" (zinc-ion batteries). We scored the relationships between the 21 new SIMs and the 14 target applications using the two models. For each new SIM, the application corresponding to the highest score was predicted as the most probable application. The model constructed via the previous approach cannot provide any prediction for the 21 new SIMs. In sharp contrast, although the model constructed via our approach has also never seen the 21 new SIMs, it can predict their future applications with high precision. Impressively, 16 of the predicted applications for the 21 new SIMs are consistent with the results in the literature [25–38] (Fig. 4). While the first application predicted for $NiBi_2Te_4$ is inconsistent with the literature [39], the application ranking second is exactly the reported application [39], i.e., "TIs", (Fig. 4). In other words, our model only fails to provide reasonable prediction results for two extremely complex SIMs, $Cu_{99}Au$ and $Os_{22}Al_{78}$ [40,41]. Notably, although $NiBi_2Te_4$ had been studied before the year 2022 [42], it has not been mentioned in the abstract and title of the paper. As discussed in the "Model construction" section, the model construction is based on the abstracts and titles; thus, $NiBi_2Te_4$ is a new material for the two models used here. The striking consistency with the subsequent reports indicates that through learning the knowledge stored in past literature, our model can well cope with SIMs that it has never seen and reasonably guide the R&D of new functional materials in a variety of subfields in materials science.

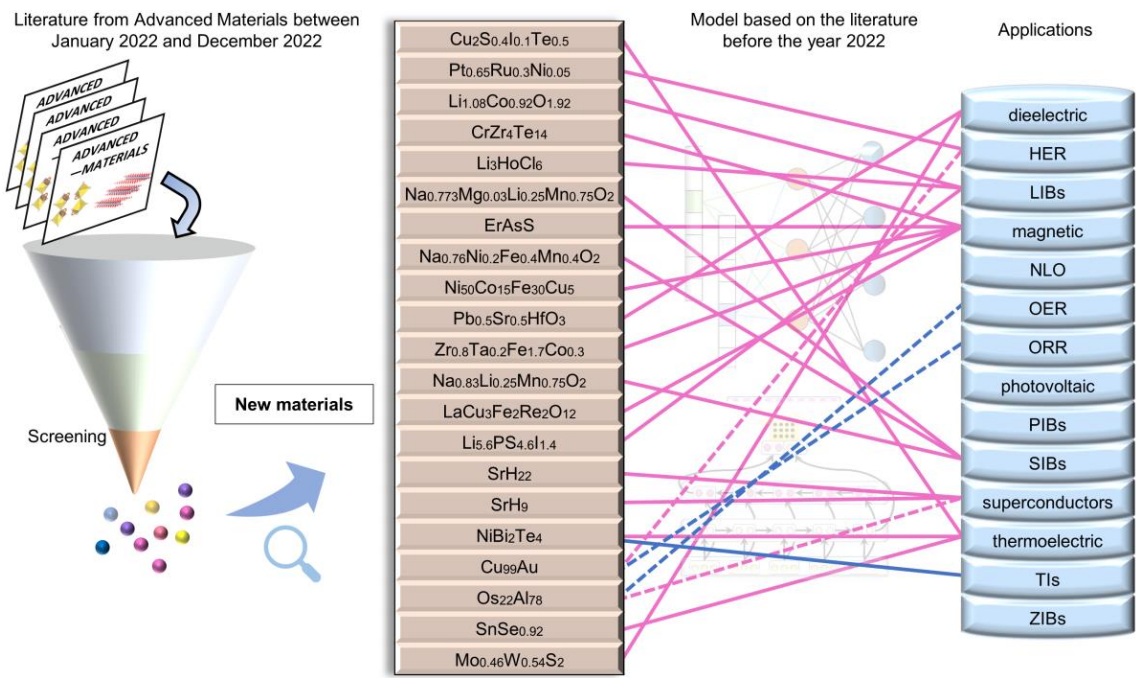

Fig. 4. Prediction for the future applications of 21 new SIMs before the year 2022. New materials before the year 2022 refer to materials that have never been reported in the abstracts and titles of the literature before the year 2022. A total of 21 new SIMs were extracted from the abstracts and titles of the literature published in Advanced Materials between January 2022 and December 2022 (Table S1). We constructed a model based on the abstracts and titles of the literature before the year 2022. Notably, since the model construction is based on the abstracts and titles of the literature before the year 2022, the model has never seen the 21 materials (that is, the 21 mate-rials are new materials for the model). The model selected the promising applications for the 21 SIMs from 14 common functional applications (8 physics-related applications and 7 chemistry-related applications). The solid line represents that the selected application that ranks first or second is consistent with the result reported in the literature. The dashed line represents that the selected application ranking first or second is inconsistent with the reported literature. The purple and blue lines represent the first and second ap-plications, respectively.

## 3. Conclusions

The model constructed via our approach can generate the word embeddings of any SIM, regardless of whether it exists in the collected literature used for model construction. HVP further confirms that by calculating cosine similarity between the word embeddings of SIMs and the target applications, the applications of both existing and SIMs can be reasonably predicted by our model. Our model is on par with the model constructed via the previous approach [18] in predicting the applications of existing SIMs (notably, the previous model can only predict the applications of existing SIMs [19]). More impressively, our model shows excellent performance in predicting the applications of new SIMs: the performance is even superior to that in the prediction for existing SIMs. It is worth highlighting that our model is the first model that can predict the future applications of any SIM. The fundamental reason for the resounding success of our approach is that it allows the generation of the word embeddings of out-of-corpus materials, which cannot be achieved by the previous approach. New SIMs extracted from the journal of Advanced Materials exemplify that our model enables a new paradigm for the R&D of new functional materials: researchers input a new SIM into a laptop and then know its promising functional applications.

## 4. Methods

### 4.1. Data collection and processing

The data used in this work was collected from materials science literature and the Materials Project [43]. The corpus was established based on the abstracts and titles of the literature from a particular year and earlier. The year 1921 was chosen as the start point for literature collection. The literature was collected through Elsevier's Scopus application programming interface (API) [44]. Natural Language Toolkit was used to segment and tokenize the corresponding abstracts and titles [45]. Meaningless punctuations and stop words ("it", "is", "that", etc.) were removed from the corpus to improve model efficiency. The tokens that were identified as valid chemical formulae by pymatgen [46] were standardized (e.g. "MoSe$_2$" was converted into "Mo1Se2"). Chemical formulae without standard regularization, such as "Bi$_{1-x}$Sb$_x$", were discarded from the corpus. The workflow for corpus establishment is presented in Fig. S3. We used the MPDataRetrieval toolkit in the matminer package [47] to extract the chemical formulae from the Materials Project. The extracted chemical formulae were standardized in the same way as described above.

*4.2. Word2vec training*

The Word2vec algorithm was implemented in genism, a python extension package [48]. The optimization of Word2vec was performed using the same hyperparameters as those used in the previous study [18], since they have proved reliable for the generation of word embeddings. 200-dimensional embeddings with a negative sampling loss of 15 were utilized in Word2vec skip-gram. The remaining key hyperparameters are as follows: we applied a learning rate of 0.01, 30 epochs, a window size of 8, and subsampling with a threshold of 0.0001. More details on the generation of word embeddings by Word2vec are presented in Supplementary Material §2 and Fig. S4.

*4.3. Word2vec training*

The standardized chemical formulae and the original word embeddings generated by Word2vec were used as input for training. The latter were represented by a matrix $R^{m \times n}$, where $m$ is the number of words, and $n$ is the dimensionality of the embedding. All characters of each formula were converted into an embedding. The corresponding character-embedding sequence of each chemical formula was processed by a forward-LSTM and a backward-LSTM, producing the final hidden vector $F_h$ and $B_h$ for the forward-LSTM and backward-LSTM, respectively. The two vectors were then input into a multilayer perceptron. Finally, RWEs were generated by $f(x) = O_T \cdot g(T_h \cdot [F_h; B_h] + b_h) + b_T$, where $x$ is the above-mentioned sequence, $O_T$, $T_h$, $b_h$ and $b_T$ are the affine transformation parameters, and $g$ represents a nonlinear elementwise function. The goal of the training is to ensure $f(x) \approx \omega_a$ by optimizing the mapping function, where $\omega_a$ represents the original word embedding. The gap between $f(x)$ and $\omega_a$ was measured by a loss function, defined as the squared Euclidean distance $\rho = ||f(x) - \omega_a||_2^2$. The key hyperparameters were set as follows: we applied 2 hidden layers, a learning rate of 0.001, and 50 epochs. The results of hyperparameter optimization are presented in Fig. S5.

**Acknowledgements**

Y. Wu and T. Liu contributed equally to this work. This work was supported by the National Natural Science Foundation of China (21825103, 22103026, 22033002, and U21A2069), the National Key Research and Development Program of China (2021YFA1500700), and the Fundamental Research Funds for the Central Universities (2020kfyXJJS050).

**Appendix A. Supplementary material**

Supplementary data to this article can be found online at (this will be filled in by the editorial staff).

**Author contributions**

Y.Z., J.W., and T.Z. conceived the original idea of this research and supervised the research project. Y.W., T.L., and H.S. demonstrated the original idea by performing theoretical analysis and computations. Y.W., T.L., and Y.Z. wrote the manuscript. J.G., K.L, and H.L discussed the results and provided valuable advices on this project.

**Data availability**

The data and code (with detailed instructions for using the code) can be obtained from GitHub (https://github.com/352worker/Reconstruct-mat2vec) or the corresponding authors upon reasonable request.

## Declaration of Competing Interest

The authors declare that they have no known competing financial interests or personal relationships that could have appeared to influence the work reported in this paper.